# STUDY OF THE USABILITY OF

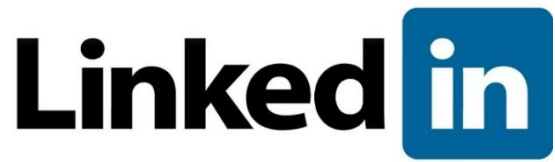

# A SOCIAL MEDIA PLATFORM MEANT TO CONNECT EMPLOYERS AND EMPLOYEES

Alessandro Ecclesie Agazzi
*Department of Computing and Informatics*
*Bournemouth University*
Poole, United Kingdom
alessandro@ecclesieagazzi.com

*Abstract—* Social network platforms have increased and become very popular in the last decade; They allow people to create an online account to then interact with others creating a complicated net of connections. LinkedIn is one of the most used Social media platforms, created and used for professional purposes. Here, indeed, the user can either apply for job positions or join professional communities to deepen his own knowledge and expertise and be always up to date in the interested field.

The primary objectives of this paper are assessing LinkedIn's usability, by using both user and expert evaluation and giving recommendations for the developer to improve this social network. This has been achieved through different steps; initially, feedbacks have been collected, via questionnaire, from direct users. Later, the usability issues, which have been underlined by users in the questionnaire, have been explored, by simulating user's problem-solving process, through Walkthrough. Finally, the overall usability of LinkedIn application has been measured by using SUS (System Usability Scale).

## A. INTRODUCTION

### I. INTRODUCING THE TOPIC

Social network platforms have increased and become very popular in the last decade (Ortiz-Ospina, 2018). They allow people to create an online account to then interact with others creating a complicated net of connections (Ortiz-Ospina, 2018). They are meant to let users keep in touch with knowns, share information and be informed about news and facts happened around the globe (Basak, Calisir 2014).

LinkedIn is one of the most used Social media platforms, created and used for professional purposes. At the present time, it counts more than 610 million members who spend, on this application, an average of 17 minutes per month, 40% of them use it daily (99firms.com, 2020). The 92% of the Fortune 500 companies use this social platform and, moreover, 57% of all the companies has a LinkedIn profile, since 2013 on (99firms.com, 2020).

Launched on the 5th of May 2003, it rapidly became the largest professional network. Here, indeed, the user can either apply for job positions or join professional communities to deepen his own knowledge and expertise and be always up to date in the interested field (Florenthal, 2015).

LinkedIn members are called "connections" and their profile pages emphasize education, employment history and skills. Moreover, it has professional network news feeds and a limited number of customisable modules (Rouse, 2020).

### II. AIM STATEMENT

The aim of this study is to assess the usability of the online social platform LinkedIn, by using both user and expert evaluation and, thus, give recommendations for the developer to improve this social network (Kiyan, 2019) (Al-Badi et al., 2013). In addition, a big attention will be given to the measure of performance of expert and non-expert users in applying for a job and interacting with the community inside the app (Al-Badi et al., 2013).

### III. LIST OF OBJECTIVES

- Identifying usability issues with task completion using LinkedIn's application
- Collecting feedbacks from direct users about LinkedIn's usability through a questionnaire
- Exploring usability issues by simulating user's problem-solving process through Walkthrough
- Measuring the overall usability of LinkedIn's application by using SUS (System Usability Scale)
- Providing recommendations to developers to improve LinkedIn's application.

## IV. RICH PICTURE

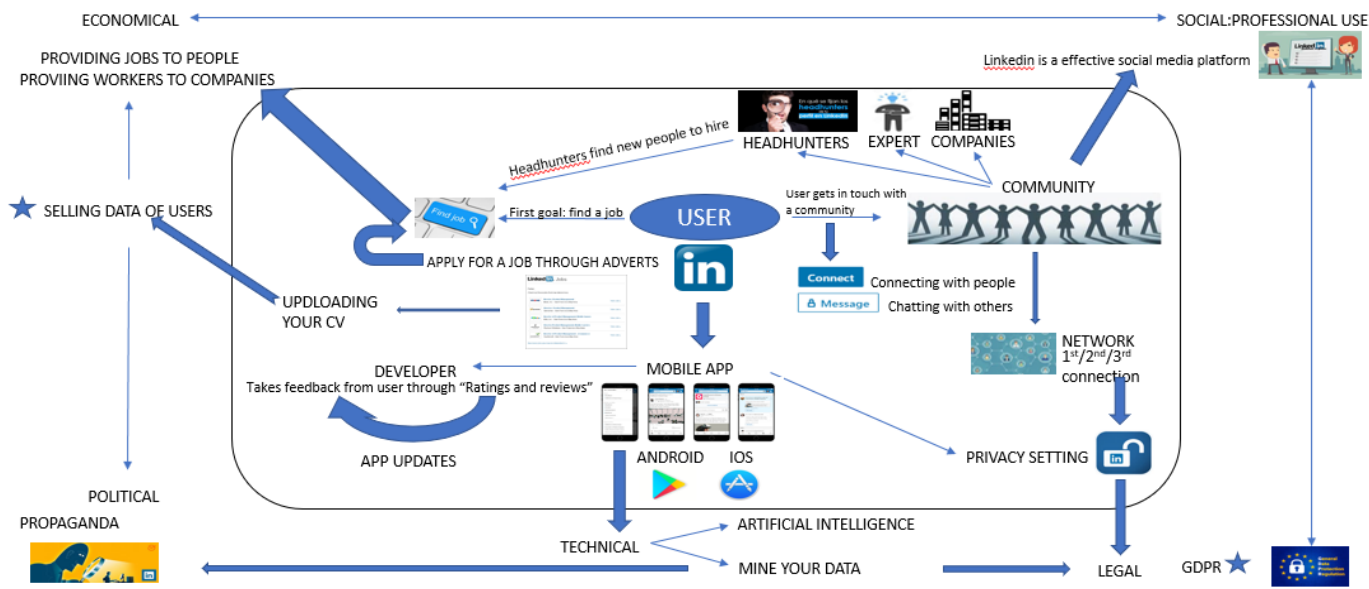

## V. DESCRIPTION OF RICH PICTURE

The above Rich picture is meant to identify the main issues of the studied case.

At the centre of the internal boundary there is the LinkedIn's user who has two main goals (Jensen, 2017):

- At the left part of the internal boundary "FINDING A JOB"
- At the right part of the internal boundary "GETTING IN TOUCH WITH A COMMUNITY"

The user first signs up on LinkedIn with the goal of getting a new job. Once registered, he can get through the adverts available on the section "jobs" and eventually apply for those he retains suitable and interesting. By doing so, he uploads his CV.

The second goal is joining different professional communities, through the function "Connect", "Message" and "Following". A community is composed by users, companies, experts and even head hunters who facilitate the first user's goal (finding a job) by looking for new employees to hire.

The network can differ according to the definition of "connection"; if you "connect" with a person, he will be considered as 1st network but a connection of him will constitute for you a 2nd network connection. As a result, a connection of a connection of a person you connected with will be considered a 3rd network connection. The "network connection" will affect what visible or not, through different steps of privacy, settable in the application's "privacy setting" (Linkedin.com, 2020).

Below the user, it has been set the section of the "mobile app" which is the logical and visual platform of LinkedIn. It is available on the main smartphones stores for different operating system: Android and IOs (Linkedin.com, 2020).

The developer uses the feedbacks, released by users in "Ratings and reviews" on the virtual stores, to create improvements through updates (Brooks, 2019).

The external boundary comprehends the following components:

- **SOCIAL**: usage of LinkedIn as "Professional social network"
- **LEGAL**: GDPR and PRIVACY SETTING, connected to the "economic issues" of selling data of users.
- **TECHNICAL**: such as artificial intelligence and the data mining meant to provide related job adverts and group communities
- **POLITICAL**: LinkedIn is sometimes used for propaganda purposes.
- **ECONOMICAL**: LinkedIn provides both jobs to people and workers to companies

### B. METHODOLOGY

## VI. QUESTIONAIRRE

"Questionnaire is a collection of questions given to different people in the same form" (Arsham, 2020).

The usability of a system can be discovered by questioning users (Al-Badi et al., 2013). This is especially efficient when dealing with the users' satisfaction or uneasiness, difficult to measure in an objective way. "Questionnaires are an indirect method" (Holzinger, 2005) because this technique gathers feedback from who uses directly the system. As a result, they are useful for studying how end users use the system and their preferred features (Al-Badi et al., 2013).

What is more, questionnaire is one of the most used techniques for gathering data from a large number of users (Wickens et al. 2003). It is a profitable way for getting much information since, indeed, questionnaires can be spread widely, leading to a big and wide amount of responses (Mathers, Fox & Hunn 2007).

All kind of questionnaires should take into consideration different notions:

➢ Questionnaires can be performed face-to-face, by phone or independently by the participants. The distinction between the method of transfer is important because it has profound effects on its design; indeed, with no interaction with the interviewer, the participants should undertake a questionnaire with a no complex filtering and simple instructions (Mathers, Fox & Hunn, 2007), like the one used for this study.

➢ Questions must be objective and not subjective in order not to spread an idea of usability as personal and subjective. The questions of this project's questionnaire were meant to be as much as objective, being verifiable at any time (Mathers, Fox & Hunn 2007).

However, sometimes the validity status of the data, received from questionnaire, can be questionable because of a lack of social cues from people answering it (Wickens et al. 2003). This method requires a sufficient number of responses to be significant (30 users being the lower limit for a study), and that it recognises fewer problems than the other methods (Al-Badi et al., 2013). In addition, the free response answers of Questionnaires tend to be short (Wickens et al. 2003). This point has been confirmed by analysing the answers of the free responses' questions of the questionnaire, used for this experiment.

For this project, questionnaire was performed in order to get information from LinkedIn users regarding their "daily" experience with this social media platform. A focus was put on the power of LinkedIn to find a new job and connect with a wide community of experts (Kiyan, 2019).

The aim is to discover which is the point of view from both expert and non-expert users who use this application.

40 people of different nationalities, 26 males and 14 females (45% of them are 26-45 years old, 25% of 46-60 years old, 22.5% between 0-25 and the remaining ones of 61 upwards), were recruited for undertaking the questionnaire which had been sent via mail through the platform Google Form.

## VII. COGNITIVE WALKTHROUGH

Cognitive walkthrough is a technique used to evaluate the usability of tasks performed by expert users. It is used to "walk through" the tasks of a system (Dalrymple, 2020) (The Interaction Design Foundation, 2020).

The usability of a system or its objectives are examined by cognitive walkthroughs. Indeed, this method is designed to assess whether a novel user carries out easily the tasks of a system (Dalrymple, 2020).

Cognitive walkthroughs is a task-specific approach; it's extremely cost-effective and also quick to perform, in comparison to different other forms of usability testing (Dalrymple, 2020).

A cognitive walkthrough begins by defining the task or tasks that the user would be expected to perform (The Interaction Design Foundation, 2020).

The 3 tasks, here analysed, are:
➢ Signing up on LinkedIn
➢ Joining a professional community
➢ Applying for a new job

Walkthrough enquiry, however, provides qualitative rather than quantitative data increasing the possibility of mistakes. In addition, it examines particular tasks instead of the overall interface (The Interaction Design Foundation, 2020).

## VIII. SYSTEM USABILITY SCALE (SUS)

This methodology provides a reliable and "quick and dirty" tool to measure the usability of a system. It is a questionnaire of 10 items with five response options (from 1 to 5) for each where 1 stands for strongly disagree and 5 strongly agree (Affairs, 2020) (Peres, Pham and Phillips, 2013).

It permits to asses a wide variety of product and services, including application, web site, hardware and software (Affairs, 2020).

SUS has became widely used with more than 1300 articles and publications using it, their benefits are:

➢ It may be used with a small number of participants with a reliable outcome (Affairs, 2020).

➢ It may precisely evaluate whether a system is usable or unusable (Peres, Pham and Phillips, 2013).

➢ It is a very easy scale to administer to participants (Affairs, 2020) (Peres, Pham and Phillips, 2013).

When this methodology is used, participants are asked to score the below 10 questions with 1 of the 5 answers which range from strongly disagree (1) to strongly agree (5) (Affairs, 2020):

➢ I would like to use LinkedIn to frequently complete task 1.

➢ I found using LinkedIn to complete task 1 unnecessarily complex.

➢ I thought using LinkedIn to complete task 1 was easy to use.

➢ I think I would need the support of a technical person to be able to use LinkedIn to complete task 1.

➢ I found the various functions of LinkedIn needed to complete task 1 were well integrated

➢ I thought there was too much inconsistency in LinkedIn to complete task 1.

➢ I would imagine that most people would learn to use LinkedIn to complete task 1 very quickly.

➢ I found LinkedIn very cumbersome to complete task 1.

➢ I felt very confident using LinkedIn to complete task 1.

➢ I needed to learn a lot of things before I could get going with completing task 1 with LinkedIn.

The interpretation of the total score may be complex. The respondent's scores for each question are converted to a new number, added together and then multiplied by 2.5 to convert the original scores of 0-40 to 0-100 (Affairs, 2020). Though the scores are 0-100, these are not percentages and should be considered only in terms of their percentile ranking (Peres, Pham and Phillips, 2013).

Based on research (Affairs, 2020), a SUS score above 68 would be considered "above average" and anything below 68 is "below average", however the best way to interpret your results involves "normalising" the scores to produce a percentile ranking (Peres, Pham and Phillips, 2013).

The LinkedIn's System Usability Scale, here studied, is based on two main tasks:

➢ APPLYING FOR A JOB task
➢ JOINING A COMMUNITY THROUG GROUP COMMUNITIES task

C. RESULTS

IX. RESULTS OF QUESTIONNAIRE

Some interesting insights were found through the answers of the questionnaire. In order to give a deep overview regarding the answers I decided to divide the insights into 3 different categories:

➢ "Some statistics of LinkedIn's usage"

The 100% of the respondents know LinkedIn and 57% of them have been using LinkedIn for more than 2 years; only 2.5% started using it less than 1 month ago. Only 4 out of 40 people use the Premium version. Finally, the most used feature in LinkedIn appeared to be "following experts in order to be always up to date".

➢ "Applying for a job" feature

30% of the respondents answered that LinkedIn's most used feature is "applying for a job".

In the free response question "What do you like most about LinkedIn?" different participants underpinned how this platform is "powerful" and "reliable" for finding a new job.

In the last "applying for a job"-related question "How would you improve LinkedIn when it comes to applying for a job?", users generally answered that they would improve the way LinkedIn suggests job adverts and the rapidity of the application process, such as adding the possibility of "easy apply" to every job advertisement.

In addition, a good improvement several users mentioned is the inclusion of a "status of the application" which would be useful to know whether an application has been received, been verifying or rejected.

➢ "Joining a professional community" feature

"Following experts in order to be always up to date" is widely the most used function with 35% of participants using this platform for this reason.

It has been interesting the analysis of the answers of the question "What do you like most about LinkedIn?"; here, indeed, 32 answers out of 40 concerned the importance of being connected to a professional community to "stay always up to date".

The 67.5% of participants found "the LinkedIn's communities helpful to enhance their knowledge and expertise".

In the last free response question "How would you improve LinkedIn when it comes to sharing knowledge and communications with your network?", many users thought "this platform has already a good system of sharing knowledge and communication within the network". Someone would suggest an improvement concerning the ease and rapidity of the communication, such as "taking out the acceptance of the message before starting a conversation".

In addition, a shared point has been the feed news; here, indeed, many participants stated how LinkedIn, sometimes, fails in suggesting posts which are not related to user's own interest.

X. RESULTS OF WALKTHROUGH

The registration process on LinkedIn is straightforward and user-friendly (Kiyan, 2019). In the walkthrough analysis only 2 issues were found:

➢ When it comes to create a new password, the further box "confirm your password" is not present. This is particularly problematic when users make mistakes in digiting their own password.

➢ The second one is related to the "reviewing your contacts" at the end of the registration process. The registration concludes only once you add at least 5 people to your network.

The second analysed task has been "connecting with other users". Some minor usability issues were found here, in particular regarding the feedbacks from the actions.

Indeed, looking for someone who does belong to the 3rd connection upwards, the "icon preview" of the desired person is absent. In addition, the "fast check" of his profile would be impossible to perform whenever that person set a "private profile".

Another usability problem, related to "connecting with other users", appears when, found the person to connect with, the user wants to send him the "connection request". Here, indeed, the "connect" button lays behind another sub menu, "…". Furthermore, the "…" menu is smaller than the nearby "message" one.

Finally, once the request is sent, the feedback, the user receives, is just a page with other suggested connections and the notification "your invitation has been sent" takes place only in a small grey box, at the top of the page.

The 3rd and last task, here analysed, is "applying for a job". Being one of the most important platforms to apply for jobs, LinkedIn has improved this feature along these years (Florenthal, 2015). Now, this system seems fluent and user-friendly (Al-Badi et al., 2013). However, there have been few aspects which require attention:

➢ Opening up the "search box" and searching for a job position leads the system to give some results. If the user wants to refine those findings with a postcode, modifiable in the filter box, the system goes back to the "general search" mixing "people", "companies" and "job adverts" results, making the research longer in terms of time and effort. What is more, once selected again the "jobs" category, LinkedIn gives many results, even those which do not belong to the desired and selected postcode.

- Once the user has applied for a job, the feedback he gets is a very small and green sentence, "You applied for this job", below a bigger "SAVE" button.

## XI. RESULTS OF SUS

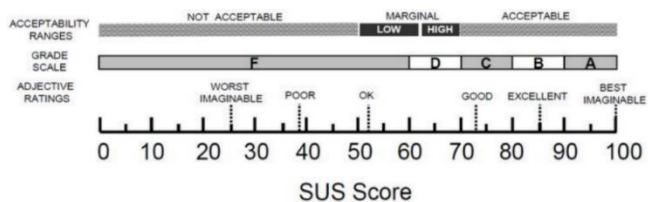

Image taken from here: (Owens, 2016)

System Usability Scale (SUS) aims to evaluate how usable and user-friendly a system is (Peres, Pham and Phillips, 2013) (Owens, 2016).

This project focuses on LinkedIn features' "applying for a job" and "joining a community".

The outcome of SUS is in line with the one of Walkthrough technique, showing how this social media platform is in general usable (Kiyan, 2019).

The scores based on 10 people of different nationality and background, but united by their comprehension and usage of LinkedIn, are 78.5 for the task "applying for a job" and 84 for "joining a community".

The acceptability range for the first task is good with a grade scale of "C" while "joining a community" gave an almost excellent acceptability range with a grade of scale "B".

### D. DISCUSSION AND LESSONS LEARNT

Although the results gathered in this experiment confirm that LinkedIn is in general a usable and user-friendly platform, especially when it comes to joining a community of experts, there are some aspects this social media would have to work on (Kiyan, 2019).

As observed during the analysis of the feedback of both expert and non-expert users, the main issue areas are the rapidity and ease of using LinkedIn (Al-Badi et al., 2013). Despite they pointed out their general satisfaction toward LinkedIn features, they brought some interesting insigths about how to improve it; what most emerged was the "introduction of the status of application", the "easy-to-apply" to every job advertisement or a more reliable system of posts' suggestion (Kiyan, 2019).

Following the set of suggestions, coming from the outcome of the adopted techniques, on how to improve the overall usage of LinkedIn that should solve the few usability issues (Al-Badi et al., 2013):

- Increasing the rapidity of how the system gives back to the users the results of post and job adverts.
- Improving the ease and rapidity of "applying for a job" by adding to all the adverts the "easy to apply" feature.
- Increasing the reliability of the application process by introducing the "status bar" which would alert the users regarding the status of their application.
- Improving the algorithm which consents to provide to users more interesting-related posts and groups.
- Fixing the box research-related issue regarding the "filtering" feature.
- Improving the registration process by both adding the "password confirmation" and removing the "review the contact you may know" step.
- Fixing the "connect" button issue by increasing its visibility and dimension.

### E. CONCLUSION AND FUTURE WORKS

Although some usability issues were found while performing the analysed tasks (applying for a job and joining a community of experts), it seems that LinkedIn is highly user-friendly (Al-Badi et al., 2013) (Kiyan, 2019). The evidence for this is the high score for both the SUS experiment (good usability for "applying for a job" feature and excellent for "joining a community of experts") and the answers in the questionnaire, which showed, in general, a users' satisfaction towards this platform and its "network's power". The overall result demonstrates that "no catastrophic usability problems" were encountered (Al-Badi et al., 2013). As a result, only those areas, which respondents pointed out as being problematic, need to improve.

Further experiments can be conducted by analysing the following problems in order to overcome the drawbacks that occurred in this study (Al-Badi et al., 2013):

- Employing a wider number of participants for both the SUS experiment and the questionnaire, in order to improve the quality of the result (Al-Badi et al., 2013).
- Employing a balanced gender ratio for both questionnaire and SUS.
- Using remote testing or employing a set of specific generated heuristic for social networking sites (Al-Badi et al., 2013).
- Replicating the same study for different social medias in order to compare each result.

# F. APPENDIX

Here below, cognitive walkthrough and SUS experiment are reported. The questionnaire has not been included because it would have taken up too much space, for more info do visit here

*COGNITIVE WALKTHROUGH*

➢ Signing up on LinkedIn

**Green: Yes**
**Red: No**

| Task 1: Signing UP | Will user know what subgoal he wants to achieve? | Will user notice that the correct action is performed? | Will user understand that the subgoal can be achieved by the action? | Does the user get feedback from the action? |
|---|---|---|---|---|
| Sub task 0: Make sure you have downloaded the app | Yes | Yes | Yes | Yes |
| Sub task 1: Select "JOIN NOW" feature | Yes | Yes | Yes | Yes |
| Sub task 2: Enter all the needed information (First name, Last Name, Email, Password) | Yes | No possibility to confirm your password | Yes | Yes |
| Sub task 3: Select "AGREE AND JOIN" | Yes | Yes | Yes | Yes |
| Sub task 4: Enter the informatione of "Profile" (I'm a student?, Most recent job title, Most recent company) | Yes | Yes | Yes | Yes |
| Sub task 5: Select next | Yes | Yes | Yes | Yes |
| Sub task 6: Enter information regarding your "area" (Country/Region, Postal Code) | Yes | Yes | Yes | Yes |
| Sub task 7: Select next | Yes | Yes | Yes | Yes |
| Sub task 8: Review the contact you may know in "community" page | No need to add people in the registration process | Yes | Yes | Yes |
| Sub task 9: Select Next to conclude your registration | Yes | Yes | Yes | Yes |

➢ Joining a professional community

**Green: Yes**
**Red: No**

| Task 2: Connecting with other users | Will user know what subgoal he wants to achieve? | Will user notice that the correct action is performed? | Will user understand that the subgoal can be achieved by the action? | Does the user get feedback from the action? |
|---|---|---|---|---|
| Sub task 0: Make sure you are registered on Linkedin | | | | |
| Sub task 1: Type the name of the person you want to connect with in "search box" | Yes | Yes | Yes | If you look for a person who does belong to the 3rd connection up, the system does not give you the "icon preview" |
| Sub task 2: Select search button | Yes | Yes | Yes | Yes |
| Sub task 3: Click on the Icon of the interested person | Yes | Yes | Yes | Yes |
| Sub task 4: Fast check the profile to be sure is the person you want to connect with | Yes | Yes | No if that profile is private | No if that profile is private |
| Sub task 5: Select "…" menu | Yes | There is no cues that going through this menu the user will find the "connect" Button | The "…" menu buttom is smaller than the "message" one | Yes |
| Sub task 6: Select the Icon "Connect" to send up the request of connection | Yes | The notification "Your invitation has been sent" is small and put at the top of the page | Yes | The notification "Your invitation has been sent" is small and put at the top of the page |
| Sub task 7: Click "x" to close the page of suggested connection | Yes | Yes | Yes | Yes |

➢ Applying for a new job

**Green: Yes**
**Red: No**

| Task 3: Applying for a job | Will user know what subgoal he wants to achieve? | Will user notice that the correct action is performed? | Will user understand that the subgoal can be achieved by the action? | Does the user get feedback from the action? |
|---|---|---|---|---|
| Sub task 0: Make sure you are registered on Linkedin | Yes | Yes | Yes | Yes |
| Sub task 1: Type the name of the company/position you want to work in/for in "search box" | Yes | Yes | Yes | Yes |
| Sub task 2: Place a tick on "Jobs" box  the refine category "Jobs" | Yes | Yes | Yes | Yes |
| Sub task 3: Select "Filter" feature | Yes | Yes | The filter button is small and in the right corner of the page thus difficult to find | Yes |
| Sub task 4: Insert postcode by typing BH13 to bring up Bournemouth options | Setting a new location makes the system go back to "people" research box | The system gives adverts of companies located somewhere else, not only those nearby the desired postcode | Yes | Yes |
| Sub task 5: Select the desired "job type"  feature in the field "Job Type" | Yes | Yes | Yes | Yes |
| Sub task 5: Select "Apply"  feature | Yes | Yes | Yes | Yes |
| Sub task 7: Check the adverts and select the preferred one | Yes | Yes | Yes | Yes |
| Sub task 8: Review selections and press on save button feature to further shortlist | Yes | Yes | Yes | Yes |
| Sub task 9: Select "Fast Apply" button | Yes | There is only a smal sentence below the SAVE button saying you applied for this job | Yes | There is only a smal sentence below the SAVE button saying you applied for this job |

*SYSTEM USABILITY SCALE (SUS)*

- APPLYING FOR A JOB task

### LinkedIn System Usability Scale

#### TASK 1 APPLYING FOR A JOB

| Participant | I would like to use Linkedin to frequently complete Task 1 | I found using Linkedin to complete Task 1 unnecessarily complex | I thought using Linkedin to complete Task 1 was easy to use | I think I would need the support of a technical person to be able to use Linkedin to complete Task 1 | I found the various functions of Linkedin needed to complete Task 1 were well integrated | I thought there was too much inconsistency in Linkedin to complete Task 1 | I would imagine that most people would learn to use Linkedin to complete Task 1 very quickly | I found Linkedin very cumbersome to complete Task 1 | I felt very confident using Linkedin to complete Task 1 | I needed to learn a lot of things before I could get going with completing Task 1 with Linkedin | TOTAL | TOTAL * 2.5 |
|---|---|---|---|---|---|---|---|---|---|---|---|---|
| 1 | 5 / 4 | 4 / 1 | 5 / 4 | 3 / 2 | 5 / 4 | 2 / 3 | 4 / 3 | 1 / 4 | 5 / 4 | 1 / 4 | 33 | 82.5 |
| 2 | 5 / 4 | 2 / 3 | 4 / 3 | 1 / 4 | 4 / 3 | 2 / 3 | 4 / 3 | 2 / 3 | 4 / 3 | 2 / 3 | 32 | 80 |
| 3 | 5 / 4 | 2 / 3 | 4 / 3 | 1 / 4 | 5 / 4 | 1 / 4 | 4 / 3 | 2 / 3 | 4 / 3 | 1 / 4 | 35 | 87.5 |
| 4 | 2 / 1 | 1 / 4 | 4 / 3 | 1 / 4 | 3 / 2 | 1 / 4 | 3 / 2 | 1 / 4 | 3 / 2 | 3 / 2 | 28 | 70 |
| 5 | 4 / 3 | 2 / 3 | 3 / 2 | 2 / 3 | 3 / 2 | 2 / 3 | 5 / 4 | 2 / 3 | 4 / 3 | 2 / 3 | 29 | 72.5 |
| 6 | 4 / 3 | 2 / 3 | 5 / 4 | 1 / 4 | 4 / 3 | 2 / 3 | 5 / 4 | 2 / 3 | 4 / 3 | 3 / 2 | 32 | 80 |
| 7 | 2 / 1 | 3 / 2 | 5 / 4 | 1 / 4 | 5 / 4 | 1 / 4 | 5 / 4 | 1 / 4 | 5 / 4 | 2 / 3 | 34 | 85 |
| 8 | 4 / 3 | 1 / 4 | 2 / 3 | 2 / 3 | 5 / 4 | 1 / 4 | 4 / 3 | 2 / 3 | 3 / 2 | 2 / 3 | 31 | 77.5 |
| 9 | 3 / 2 | 1 / 4 | 4 / 3 | 1 / 4 | 4 / 3 | 1 / 4 | 5 / 4 | 1 / 4 | 3 / 2 | 1 / 4 | 34 | 85 |
| 10 | 5 / 4 | 2 / 3 | 5 / 4 | 1 / 4 | 4 / 3 | 1 / 4 | 5 / 4 | 1 / 4 | 4 / 3 | 1 / 4 | 37 | 92.5 |

SUS Score: 78.5  Adjective Rating Scale:

- JOINING A COMMUNITY THROUG GROUP COMMUNITIES task

#### TASK 2 JOINING A COMMUNITY THROUG GROUP COMMUNITIES

| Participant | I would like to use Linkedin to frequently complete Task 1 | I found using Linkedin to complete Task 1 unnecessarily complex | I thought using Linkedin to complete Task 1 was easy to use | I think I would need the support of a technical person to be able to use Linkedin to complete Task 1 | I found the various functions of Linkedin needed to complete Task 1 were well integrated | I thought there was too much inconsistency in Linkedin to complete Task 1 | I would imagine that most people would learn to use Linkedin to complete Task 1 very quickly | I found Linkedin very cumbersome to complete Task 1 | I felt very confident using Linkedin to complete Task 1 | I needed to learn a lot of things before I could get going with completing Task 1 with Linkedin | TOTAL | TOTAL * 2.5 |
|---|---|---|---|---|---|---|---|---|---|---|---|---|
| 1 | 5 / 4 | 1 / 4 | 5 / 4 | 1 / 4 | 5 / 4 | 1 / 4 | 4 / 3 | 1 / 4 | 5 / 4 | 1 / 4 | 39 | 97.5 |
| 2 | 4 / 3 | 2 / 3 | 5 / 4 | 2 / 3 | 3 / 2 | 2 / 3 | 3 / 2 | 3 / 2 | 4 / 3 | 1 / 4 | 29 | 72.5 |
| 3 | 5 / 4 | 1 / 4 | 5 / 4 | 2 / 3 | 5 / 4 | 3 / 2 | 5 / 4 | 2 / 3 | 5 / 4 | 2 / 3 | 35 | 87.5 |
| 4 | 5 / 4 | 1 / 4 | 5 / 4 | 1 / 4 | 4 / 3 | 2 / 3 | 4 / 3 | 2 / 3 | 4 / 3 | 2 / 3 | 35 | 87.5 |
| 5 | 5 / 4 | 2 / 3 | 5 / 4 | 1 / 4 | 4 / 3 | 4 / 1 | 3 / 2 | 1 / 4 | 2 / 1 | 1 / 4 | 30 | 75 |
| 6 | 3 / 2 | 3 / 2 | 3 / 2 | 3 / 2 | 5 / 4 | 3 / 2 | 5 / 4 | 2 / 3 | 2 / 1 | 3 / 2 | 24 | 60 |
| 7 | 5 / 4 | 1 / 4 | 4 / 3 | 1 / 4 | 5 / 4 | 3 / 2 | 4 / 3 | 1 / 4 | 4 / 3 | 2 / 3 | 34 | 85 |
| 8 | 5 / 4 | 2 / 3 | 2 / 1 | 2 / 3 | 4 / 3 | 2 / 3 | 5 / 4 | 2 / 3 | 5 / 4 | 2 / 3 | 31 | 77.5 |
| 9 | 4 / 3 | 2 / 3 | 4 / 3 | 1 / 4 | 4 / 3 | 2 / 3 | 5 / 4 | 3 / 2 | 3 / 2 | 2 / 3 | 30 | 75 |
| 10 | 4 / 3 | 1 / 4 | 4 / 3 | 1 / 4 | 4 / 3 | 1 / 4 | 3 / 2 | 1 / 4 | 5 / 4 | 2 / 3 | 34 | 85 |

SUS Score: 84.0  Adjective Rating Scale: